\documentclass[reprint,preprintnumbers,prd,nofootinbib]{revtex4-1}

\usepackage{amsfonts}
\usepackage{amsmath}
\usepackage{array}
\usepackage{braket}
\usepackage{epstopdf}
\usepackage{graphicx}
\usepackage{hepparticles}
\usepackage{hepnicenames}
\usepackage{hepunits}
\usepackage{hyperref}
\usepackage[%
    utf8
]{inputenc}
\usepackage{slashed}
\usepackage{subfigure}
\usepackage{placeins}
\usepackage[%
    normalem
]{ulem}
\usepackage[%
    usenames,
    svgnames,
    dvipsnames
]{xcolor}

\newcommand{\order}[1]{\mathcal{O}\left({#1}\right)}
\newcommand{\refapp}[1]{appendix~\ref{app:#1}}
\newcommand{\refeq}[1]{eq.~(\ref{eq:#1})}
\newcommand{\reffig}[1]{figure~\ref{fig:#1}}
\newcommand{\refsec}[1]{section~\ref{sec:#1}}
\newcommand{\reftab}[1]{table~\ref{tab:#1}}

\renewcommand{\theta}{\vartheta}

\newcommand{\para}{\parallel}

\newcommand{\dd}[2][]{{\mathrm{d}^{#1}}#2\,}
\newcommand{\mLambdaB}{m_{\Lambda_b}}
\newcommand{\mLambda}{m_{\Lambda}}

\DeclareMathOperator*{\argmax}{arg\,max}
\newcommand{\wilson}[1]{\mathcal{C}_{#1}}
\newcommand{\op}[1]{\mathcal{O}_{#1}}
\newcommand{\la}{\langle}
\newcommand{\ra}{\rangle}

\begin{document}

\allowdisplaybreaks

\preprint{EOS-2016-02, RBRC-1168, ZU-TH-7/16}
\title{Using $\Lambda_b\to \Lambda\mu^+\mu^-$ data within a Bayesian analysis of $|\Delta B| = |\Delta S| = 1$ decays}
\author{Stefan Meinel}
\email{smeinel@email.arizona.edu}
\affiliation{Department of Physics, University of Arizona, Tucson, AZ 85721, USA}
\affiliation{RIKEN BNL Research Center, Brookhaven National Laboratory, Upton, NY 11973, USA}
\author{Danny van Dyk}
\email{dvandyk@physik.uzh.ch}
\affiliation{Physik-Institut, Universit\"at Z\"urich, Winterthurer Strasse 190, 8057 Z\"urich, Switzerland}

\begin{abstract}
    We study the impact of including the baryonic decay $\Lambda_b\to \Lambda(\to p\, \pi^-)\mu^+\mu^-$
    in a Bayesian analysis of $|\Delta B | = |\Delta S| = 1$ transitions. We perform fits of the
    Wilson coefficients $\wilson{9}$, $\wilson{9'}$, $\wilson{10}$ and $\wilson{10'}$, in addition
    to the relevant nuisance parameters. Our analysis combines
    data for the differential branching fraction and three angular observables of
    $\Lambda_b\to \Lambda(\to p\, \pi^-)\mu^+\mu^-$ with data for the branching ratios of
    $B_s \to \mu^+\mu^-$ and inclusive $b \to s\ell^+\ell^-$ decays.
    Newly available precise lattice QCD results for the full set of $\Lambda_b \to \Lambda$ form factors are used
    to evaluate the observables of the baryonic decay.
    Our fits prefer shifts to $\wilson{9}$ that are opposite in sign compared
    to those found in global fits of only mesonic decays,
    and the posterior odds show no evidence of physics beyond the Standard Model. We investigate a possible hadronic origin
    of the observed tensions between theory and experiment.
\end{abstract}

\maketitle

\section{Introduction}
\label{sec:intro}

The tensions between theory and experiment for $P'_5$ \cite{Aaij:2013qta, Aaij:2015oid}, one of the angular
observables in the kinematical distribution of the decay $\bar{B}\to
\bar{K}^*(\to \bar{K}\pi)\mu^+\mu^-$ \cite{DescotesGenon:2012zf}, have sparked
much interest in the determination of the short-distance couplings in
flavor-changing neutral currents of the form $b\to s\ell^+\ell^-$. Several
competing global analyses have been published \cite{%
Descotes-Genon:2013wba,Altmannshofer:2013foa,Beaujean:2013soa,Horgan:2013pva,Hurth:2013ssa,%
Altmannshofer:2014rta,Hurth:2014vma,Beaujean:2015gba, Du:2015tda, Descotes-Genon:2015uva,%
Hurth:2016fbr} that use the available
data on such rare decays of $\bar{B}$ mesons to various degrees, and most of these analyses
find that a negative shift in the Wilson coefficient $\wilson{9}$
improves the agreement with the data. However, it remains unclear whether this effect is caused by physics
beyond the Standard Model, or merely by uncontrolled hadronic contributions.

None of the published analyses include the first measurements of angular observables of the
baryonic rare decay $\Lambda_b\to \Lambda(\to p\,\pi^-)\mu^+\mu^-$
\cite{Aaij:2015xza}, which offers complementary constraints compared to the
commonly used mesonic channels. A recent lattice QCD calculation of the relevant
$\Lambda_b \to \Lambda$ form factors \cite{Detmold:2016pkz} enables us to
evaluate the $\Lambda_b\to \Lambda(\to p\,\pi^-)\mu^+\mu^-$ observables with
high precision. The purpose of this work is thus to study the constraining
power of the $b$-baryon decay in a global analysis of $|\Delta B| = |\Delta S| = 1$ decays.

Our article is structured as follows: First, we briefly describe our framework,
define our fit models and review the observations that enter our likelihood
function in \refsec{framework}. We then present our results for each of the
fit models in \refsec{results}, and further discuss the implications in \refsec{summary}.
Appendix \ref{app:subleading-terms} describes the subleading corrections to the OPE at low hadronic recoil;
appendix \ref{app:predictions} gives posterior-predictive distributions for the full
set of $\Lambda_b\to \Lambda(\to p\,\pi^-)\mu^+\mu^-$ angular observables, and appendix \ref{app:lambdabonly}
contains additional fit results using only the data for the baryonic decay.

\section{Framework}
\label{sec:framework}

We work in the usual effective field theory for flavor-changing neutral $b\to
s\lbrace \gamma, \ell^+\ell^-\rbrace$ transitions; see e.g. \cite{Bobeth:2012vn}.
Its Hamiltonian reads
\begin{equation}
\begin{aligned}
  \label{eq:Heff}
  {\cal{H}}_{\rm eff}
  & = - \frac{4\, G_F}{\sqrt{2}}  V_{tb}^{} V_{ts}^* \,\frac{\alpha_e}{4 \pi}\,
       \sum_i \wilson{i}(\mu)  \op{i}\\
  & + \order{V_{ub} V_{us}^*} + \text{h.c.}\,,
\end{aligned}
\end{equation}
where $\wilson{i}(\mu)$ denotes the Wilson coefficients at the renormalization
scale $\mu$, and $\op{i}$ denotes a basis of field operators. The most relevant
operators are
\begin{equation}
\label{eq:SM:ops}
\begin{aligned}
    \op{7(7')} & = \frac{m_b}{e}\!\left[\bar{s} \sigma^{\mu\nu} P_{R(L)} b\right] F_{\mu\nu}\,,\\
    \op{9(9')} & = \left[\bar{s} \gamma_\mu P_{L(R)} b\right]\!\left[\bar{\ell} \gamma^\mu \ell\right]\,,\\
    \op{10(10')} & = \left[\bar{s} \gamma_\mu P_{L(R)} b\right]\!\left[\bar{\ell} \gamma^\mu \gamma_5 \ell\right]\,,
\end{aligned}
\end{equation}
where a primed index indicates a flip of the quarks' chiralities with respect to the
unprimed, Standard Model(SM)-like operator. Further four-quark operators
$\op{i} \sim \left[\bar{s} \Gamma_i b\right]\,\left[\bar{q}\Gamma_i^\prime q\right]$, $i=1,\dots,6$
as well as the chromomagnetic operator $\op{8}$ contribute to the transition amplitudes via
hadronic matrix elements of two-point correlators with the quark electromagnetic current. These
contributions are taken into account in the numerical evaluation of the $b\to s \ell^+\ell^-$ observables
via process- and $q^2$-dependent shifts of the effective Wilson coefficients $\wilson{9,\lambda}$
and $\wilson{7,\lambda}$ that enter into the various transversity amplitudes.
The expressions relevant to this work
can be taken from Refs.~\cite{Beylich:2011aq, Boer:2014kda}
(for $\Lambda_b\to \Lambda\ell^+\ell^-$ at high $q^2$) and
Ref.~\cite{Huber:2005ig} (for $B\to X_s\ell^+\ell^-$ at low $q^2$). For definiteness, we fix $\mu = 4.2\,\GeV$
in our fits.

\subsection{Fit Models}

For the purpose of our analysis we define three fit scenarios, labeled
``SM($\nu$-only)'', ``$(9,10)$'',  and ``$(9,9',10,10')$'', respectively:
\begin{equation}
\label{eq:scenarios}
\begin{aligned}
    \text{SM($\nu$-only)} & : \begin{cases}
        \wilson{7,9,10} & \text{SM values}\\
        \wilson{7',9',10'} & \text{SM values}\\
        \vec\nu & \text{free floating}
    \end{cases}\,,\\
    (9,10) & : \begin{cases}
        \wilson{9}         & \in [-4,+9]\\
        \wilson{10}        & \in [-6,-2]\\
        \wilson{7,7',9',10'} & \text{SM values}\\
        \vec\nu            & \text{free floating}
    \end{cases}\,,\\
    (9,9',10,10') & : \begin{cases}
        \wilson{9,9',10,10'} & \in [-8,+8]\\
        \wilson{7,7'}        & \text{SM values}\\
        \vec\nu              & \text{free floating}
    \end{cases}\,,
\end{aligned}
\end{equation}
where the parameters of interest are $\vec\theta = (\wilson{9}, \wilson{10})$ or $\vec\theta = (\wilson{9}, \wilson{9'}, \wilson{10}, \wilson{10'})$,
and where the nuisance parameters $\vec{\nu}$ account for theoretical uncertainties in the
computation of the observables\footnote{%
    Note that our fit models are lepton-flavor-universal, and therefore
    cannot account for the present measurement of $R_K$ \cite{Aaij:2014ora}.
}. We obtain the posterior density for a given model $M$,
$P(\vec{x}\,|\, M, \text{data})$, using Bayes' theorem
\begin{equation}
    \label{eq:bayes-theorem}
    P(\vec x\,|\,M, \text{data})
    = \frac{P(\text{data}\,|\,\vec x, M) P_0(\vec x\,|\,M)}
    {P(\text{data}\,|\,M)}\,.
\end{equation}
In the above, $\vec{x} \equiv (\vec{\theta}, \vec{\nu})$, $P_0(\vec x, M)$ is
the prior density, and $P(\text{data}\,|\,\vec x, M)$ denotes the product of
the experimental likelihoods. The prior density factorizes,
\begin{equation}
    P_0(\vec x\,|\, M) \equiv P_0(\vec \theta\,|\, M) P_0(\vec \nu\,|\, M)\,,
\end{equation}
into the prior for the parameters of interest, which is multivariate uniform (see \refeq{scenarios}), and
the informative (i.e., non-uniform) priors for the nuisance parameters.
The normalization on the right-hand side of \refeq{bayes-theorem},
\begin{equation}
    \label{eq:evidence}
    P(\text{data}\,|\,M) \equiv \int_{V(M)} \dd{\vec x} \,P(\text{data}\,|\,\vec x, M) P_0(\vec x\,|\,M)\,,
\end{equation}
is the total evidence of the data given the model $M$. We will refer to it as the
\emph{local evidence} whenever we restrict the integration hypervolume $V(M)$
to a subset of the support of $P_0(\vec x, M)$.
The parameter point $\vec{x}^{\,*} \equiv (\vec{\theta}^{\,*},
\vec{\nu}^{\,*})$ maximizes the posterior,
\begin{equation}
    \vec{x}^{\,*} = \argmax_{x} P(\vec x\,|\,M,\text{data})\,,
\end{equation}
and is referred to as the best-fit point. For the purpose of calculating the
goodness of fit, we then compute
\begin{equation}
    \chi^2 \equiv -2\ln P(\text{data}\,|\,\vec{x}^{\,*}, M)\,.
\end{equation}
Since all measurements enter the likelihood as univariate Gaussians, we define
their individual \emph{pull} values as
\begin{equation}
    \operatorname{pull}_i \equiv \frac{O - O(\vec{x}^{\,*})}{\sigma}\,,
\end{equation}
in which $O \pm \sigma$ corresponds to the experimental results, and $O(\vec{x}^{\,*})$
denotes the theory prediction at the best-fit point.

In order to compare pairs of fit models, we employ the notion of posterior odds.
The odds of model $M_1$ over model $M_2$ are defined as
\begin{equation}
    \frac{P(M_1\,|\,\text{data})}{P(M_2\,|\,\text{data})} = \frac{P(\text{data}\,|\,M_1)}{P(\text{data}\,|\,M_2)} \frac{P_0(M_1)}{P_0(M_2)}\,.
\end{equation}
In the above, $P_0(M)$ denotes a model prior. The latter can
be obtained from, e.g., independent fits. In the absence of such results
and following standard practice, we use
identical priors for all our models: $P_0(M)\equiv 1 \,\forall\, M$.

Our statistical approach closely follows the one used in
Refs.~\cite{Beaujean:2012uj,Beaujean:2013soa}.  The calculation of all observables
(listed in the following subsection), and the statistical procedures are
carried out through use of the EOS software \cite{EOS}, which implements a
Monte Carlo algorithm as described in Ref.~\cite{Beaujean:2013}.

\subsection{Inputs}

Our fits take into account the following observables:
\begin{enumerate}
    \item The main task is the inclusion of the branching ratio of
        $\Lambda_b\to \Lambda(\to p\,\pi^-)\mu^+\mu^-$ decays, as well as three further
        observables that arise from the angular distribution \cite{Boer:2014kda}:
        $F_0$, the rate of longitudinally-polarized lepton pairs, as well as the
        leptonic and the hadronic forward-backward asymmetries $A_\text{FB}^\ell$
        and $A_\text{FB}^\Lambda$. The theory of QCD factorization at low $q^2$ \cite{Beneke:2001at}
        is not yet fully developed for the baryonic decay (see Ref.~\cite{Wang:2015ndk} for a recent
        discussion), and we therefore restrict
        our analysis to the high-$q^2$ region, where the usual low-recoil OPE \cite{Grinstein:2004vb, Beylich:2011aq}
        is applicable. This restricts our use to observables that are integrated over the entire low recoil
        region, $15\,\GeV^2 \leq q^2 \leq  20\,\GeV^2 \simeq q^2_\text{max}$. We denote
        the binning in this range as $\langle \cdot \rangle_{15,20}$.

        The LHCb collaboration has published an analysis of both the branching ratio
        and the three aforementioned angular observables \cite{Aaij:2015xza}, which
        are all included in our likelihood.
        The CDF collaboration had previously reported \cite{Aaltonen:2011qs} the first
        observation of this decay, and performed a measurement of its branching ratio.
        However, the CDF analysis is based on only a small number of $24\pm 5$ signal candidates
        in the entire phase space; the uncertainty of the branching ratio in the low recoil region
        is accordingly large. The CDF result is compatible with the LHCb result,
        but with approximately three times larger uncertainty. We therefore do not
        include the CDF measurement in our fits.

    \item We denote the time-integrated branching ratio of the decay $B_s \to \mu^+\mu^-$
        as $\int \dd{\tau} \mathcal{B}(\tau)$ \cite{DeBruyn:2012wk}. Our likelihood
        includes the recent results from a combined analysis of the CMS and LHCb
        collaborations \cite{CMS:2014xfa}.
        All of our fit models, as specified in \refeq{scenarios}, imply $A^{\mu\mu}_{\Delta \Gamma_s} = 1$ \cite{DeBruyn:2012wk}.

    \item From the inclusive decay $B\to X_s\ell^+\ell^-$ we use the branching ratio,
        integrated over the range of dilepton mass square $1\,\GeV^2 \leq q^2 \leq 6\,\GeV^2$;
        denoted as $\langle \mathcal{B}\rangle_{1,6}$.
        The likelihood includes the measurements by the BaBar \cite{Lees:2013nxa} and the
        Belle \cite{Iwasaki:2005sy} \footnote{%
            In absence of a measurement for the $\mu^+\mu^-$ final state, we use the Belle result
            for a mixture of $\mu^+\mu^-$ and $e^+e^-$ final states, assuming lepton universality.
        } collaborations.
\end{enumerate}
A summary of nuisance parameters, their association with specific
observables, and their respective priors that enter our analyses is shown in
\reftab{nuisance-parameters}.

The ten $\Lambda_b \to \Lambda$ form factors in the helicity basis are parametrized using simplified $z$ expansions
\cite{Bourrely:2008za} of the form
\begin{equation}
 f(q^2) = \frac{1}{1-q^2/m_{{\rm pole},f}^2} \sum_{k=0}^{k_{\rm max}} a_{f,k}\:[z(q^2)]^k.
\end{equation}
The prior distribution of the parameters $\{a_{f,k}\}$ is a multivariate
Gaussian given by the lattice QCD calculation of
Ref.~\cite{Detmold:2016pkz} (the definition of $z$ and the values of the pole masses, $m_{{\rm pole},f}$, are
also given in Ref.~\cite{Detmold:2016pkz}).  Note that Ref.~\cite{Detmold:2016pkz} provides
two sets of form factor parameters: the ``nominal parameters'' with $k_{\rm
max}=1$, which are used to evaluate central values and statistical
uncertainties, and the ``higher-order parameters'' with $k_{\rm max}=2$, which
are used in combination with the nominal parameters to evaluate systematic
uncertainties according to Eqs.~(50)-(56) of Ref.~\cite{Detmold:2016pkz}. Since
a Bayesian fit requires a single fit model, we
follow a simplified approach in this work. We use $k_{\rm max}=2$ throughout,
but set the central values of $a_{f,0}$ and $a_{f,1}$ equal to the nominal
values and set the central values of $a_{f,2}$ to zero. We then compute the
\emph{total} (statistical plus systematic) covariance matrix of the parameters
$\{a_{f,0}, a_{f,1}, a_{f,2} \}$ according to Eq.~(56) of
Ref.~\cite{Detmold:2016pkz}, and use this total covariance matrix in our prior
distribution. In the high-$q^2$ region considered here, this simplified
procedure accurately reproduces the total covariances of all form factors and
observables as computed using the original method \cite{Detmold:2016pkz}.
\footnote{This
is not the case in the low-$q^2$ region (which is not used here). At low $q^2$,
the statistical and systematic uncertainties in the form factors are larger due to the
absence of lattice data points in that region. Consequently, deviations from the quadratic
approximation in Gaussian error propagation are larger, and the resulting estimates
depend on the order of the steps taken.
}

\begin{table}
\begin{center}
\renewcommand{\arraystretch}{1.4}
\begin{tabular}{c|ccc}
  Quantity & Prior & Unit & Reference\\
\hline
  \multicolumn{4}{c}{CKM Wolfenstein parameters}\\
\hline
  $A$                            &  $0.806 \pm 0.020$      &  ---      &  \cite{Bona:2006ah}\\
  $\lambda$                      &  $0.2253 \pm 0.0006$    &  ---      &  \cite{Bona:2006ah}\\
  $\bar{\rho}$                   &  $0.132 \pm 0.049$      &  ---      &  \cite{Bona:2006ah}\\
  $A$                            &  $0.369 \pm 0.050$      &  ---      &  \cite{Bona:2006ah}\\
\hline
  \multicolumn{4}{c}{Quark masses}\\
\hline
  $\overline{m}_c(m_c)$          &  $1.275 \pm 0.025$      &  $\GeV$   &  \cite{Agashe:2014kda}\\
  $\overline{m}_b(m_b)$          &  $4.18 \pm 0.03$        &  $\GeV$   &  \cite{Agashe:2014kda}\\
\hline
  \multicolumn{4}{c}{HQE parameters}\\
\hline
  $\mu^2_{\pi}(1\, \mbox{GeV})$  &  $0.45 \pm 0.10$        &  $\GeV^2$ &  \cite{Uraltsev:2001ih}\\
  $\mu^2_{G}(1\, \mbox{GeV})$    &  $0.35^{+0.03}_{-0.02}$ &  $\GeV^2$ &  \cite{Uraltsev:2001ih}\\
\hline
  \multicolumn{4}{c}{$B_s$ decay constant}
\\
\hline
  $f_{B_s}$                      &  $227.7 \pm 4.5$        &  $\MeV$   &  \cite{Laiho:2009eu,Bazavov:2011aa,McNeile:2011ng, Na:2012kp}\\
\hline
  \multicolumn{4}{c}{$\Lambda\to p\,\pi^-$ decay parameter}
\\
\hline
  $\alpha$                      &  $0.642 \pm 0.013$        &  ---   &  \cite{Agashe:2014kda} \\
\end{tabular}
\renewcommand{\arraystretch}{1.0}
\caption{%
    \label{tab:nuisance-parameters}
    Prior distributions of selected nuisance parameters: Cabibbo-Kobayashi-Maskawa (CKM) parameters, quark
    masses, hadronic matrix elements entering the inclusive and the
    exclusive leptonic decays, and $\Lambda\to p\,\pi^-$ parity-violating decay parameter.
    For the CKM parameters, we use the results of a Bayesian analysis
    of only tree-level decays, which was performed by the UTfit Collaboration in 2013 \cite{Bona:2006ah}.
    All distributions are Gaussian, with the
    exception of $\mu^2_{G}(1\, \mbox{GeV})$. The latter follows a LogGamma
    distribution whose additional parameter allows us to faithfully reproduce the asymmetric
    uncertainty interval as given in \cite{Uraltsev:2001ih}. The prior distribution for the $\Lambda_b\to \Lambda$ form
    factors is a multivariate Gaussian with inputs directly taken from the
    lattice QCD calculation in Ref.~\cite{Detmold:2016pkz}; see the text
    for details.
}
\end{center}
\end{table}

\section{Results}
\label{sec:results}

In the following subsections we will summarize our findings
for each of the fit scenarios through
\begin{itemize}
    \item the value of $\vec{\theta}^{\,*}$, the best-fit point for the
        parameters of interest (if applicable);
    \item a summary of $\vec{\nu}^{\,*}$, the nuisance parameters at the
        best-fit point, as well as a summary of the 1D-marginalizd densities
        of the posterior for all components of $\vec{\nu}$;
    \item a $\chi^2$ value and its associated $p$-value: our a-priori threshold
        for an acceptable fit is $p \geq 0.03$;
    \item a description of a hypercube that includes the global mode of
        the posterior, and its associated local evidence;
    \item and a summary of the 1D-marginalized posteriors of the parameters
        of interest for the local solution with the largest local evidence.
\end{itemize}
A short summary of the goodness-of-fit quantities for each of the scenarios
is shown in \reftab{goodness-of-fit}\,.

\begin{table}[t]
\newcolumntype{C}[1]{>{\centering\let\newline\\\arraybackslash\hspace{0pt}}m{#1}}
\renewcommand{\arraystretch}{1.4}
\begin{tabular}{c|C{.11\textwidth}|C{.11\textwidth}|C{.11\textwidth}}
    ~                                            & \multicolumn{3}{c}{Pull value [$\sigma$]}                                \\
    Constraint                                   & SM($\nu$-only)         & ~$(9,10)$~             & $(9,9',10,10')$        \\
    \hline
    \multicolumn{4}{c}{$\Lambda_b\to \Lambda\mu^+\mu^-$}                                                                    \\
    \hline
    $\la \mathcal{B}\ra_{15,20}$                 & $+0.86$                & $-0.17$                & $-0.08$                \\
    $\la F_0\ra_{15,20}$                         & $+1.41$                & $+1.41$                & $+1.41$                \\
    $\la A_\text{FB}^\ell\ra_{15,20}$            & $+3.13$                & $+2.60$                & $+0.72$                \\
    $\la A_\text{FB}^\Lambda\ra_{15,20}$         & $-0.26$                & $-0.24$                & $-1.08$                \\
    \hline
    \multicolumn{4}{c}{$\bar{B}_s\to \mu^+\mu^-$}                                                                             \\
    \hline
    $\int \dd{\tau} \mathcal{B}(\tau)$           & $-0.72$                & $+0.75$                & $+0.37$                \\
    \hline
    \multicolumn{4}{c}{$\bar{B}\to X_s\ell^+\ell^-$}                                                                          \\
    \hline
    $\la \mathcal{B}\ra_{1,6}$ (BaBar)           & $+0.47$                & $-0.26$                & $-0.10$                \\
    $\la \mathcal{B}\ra_{1,6}$ (Belle)           & $+0.17$                & $-0.35$                & $-0.24$                \\
    \hline
    ~                                            & \multicolumn{3}{c}{$\chi^2$ at best-fit point}                           \\
    \hline
    ~                                            & $13.40$                & $9.60$                 & $3.87$                 \\
\end{tabular}
\renewcommand{\arraystretch}{1.0}
\caption{\label{tab:goodness-of-fit}
    Pull values for the individual experimental constraints within each of the
    fit scenarios at the respective best-fit point. The last line gives the total $\chi^2$ value
    at the respective best-fit point.
}
\end{table}

\subsection{Scenario SM($\nu$-only)}
\label{sec:scenario-SMnu}

This scenario does not feature any parameters of interest, and thus only probes
the goodness of fit between the theory predictions in the SM and the data. We find that both
the best-fit point $\vec{\nu}^{\,*}$ and the 1D-marginalized posterior
densities correspond excellently to the prior density for each of the $35$ nuisance parameters. 
Overall, we find $\chi^2 = 13.40$ for $7$ degrees of freedom (d.o.f.). This translates
to a $p$-value of $0.06$, assuming a Gaussian likelihood. Since this value is larger than our a-priori
threshold for the $p$-value, we accept this fit. We obtain the global
evidence as $P(\text{data} \,|\, \text{SM($\nu$-only)}) = (1.1469 \pm 0.0003)\cdot 10^{18}$,
where the error is only statistical in nature\footnote{Large
numbers for the evidence are not worrisome. They are driven by the integration
of the likelihood as a function of the model parameters over the model
parameters. As such, they are meaningful for comparison of fits as long as the
fits share the same likelihood.  Providing the evidence as part of our analysis
allows other researchers to make their own conclusions, and to produce their
own Bayes factors for model comparisons.
}. The individual pull values for this scenario
are listed in the left column of \reftab{goodness-of-fit}. We draw attention to the observable
$\langle A_\text{FB}^\ell\rangle_{15,20}$, whose pull is the only pull to exceed $3\sigma$; all
other pulls are smaller than $2\sigma$.

\subsection{Scenario (9,10)}
\label{sec:scenario-SM}

\begin{figure}[t]
\begin{tabular}{c}
    \includegraphics[width=.4\textwidth]{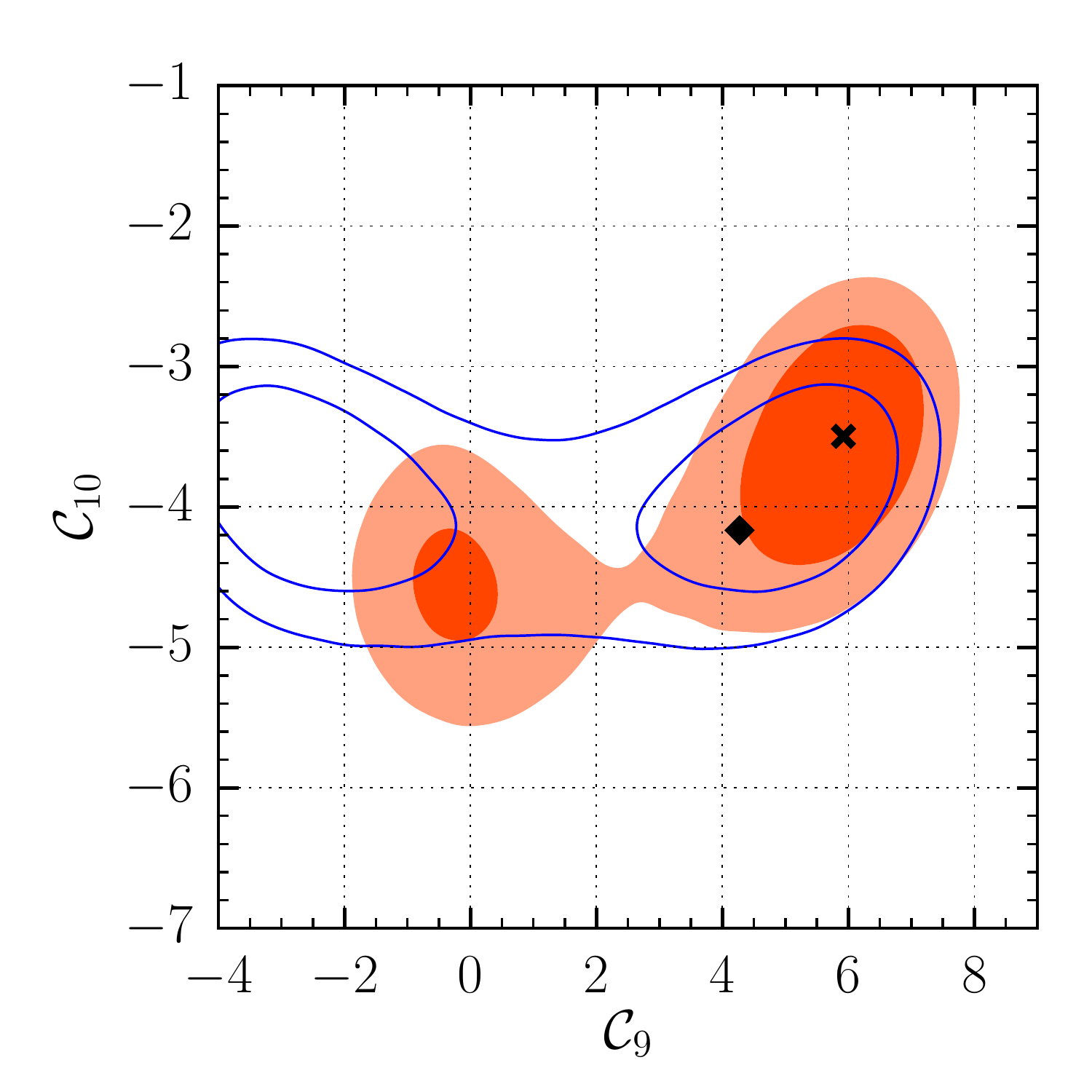}
\end{tabular}
\caption{\label{fig:posteriors-SM}
    The 2D-marginalised posterior in the $\wilson{9}$-$\wilson{10}$ plane.
    To demonstrate the impact of including the baryonic decay in the analysis,
    we show the results from a fit to the $\bar{B}\to X_s\ell^+\ell^-$ and $\bar{B}_s\to\mu^+\mu^-$ branching ratios only
    (blue lines) and from the full fit including also the $\Lambda_b\to \Lambda(\to p\,\pi^-)\mu^+\mu^-$ observables (orange-red areas).
    The SM point is marked with a diamond shape, while the best-fit point from the full fit
    is marked with a black cross. The contours correspond to $68\%$ (inner contours)
    and $95\%$ (outer contours) of probability for the respective 2D-marginalised posteriors.
}
\end{figure}

In this scenario we fit $2$ parameters of interest in addition to the parameters of SM($\nu$-only).
The $\theta$ components of the best-fit point read
\begin{equation}
    \vec{\theta}^{\,*}\!:\!(\wilson{9}, \wilson{10}) = ( 5.92, -3.50 )\,.
\end{equation}
The 2D marginalization onto these two parameter is shown in \reffig{posteriors-SM}.
As before for the model SM($\nu$-only), we find that also in this fit model the nuisance parameters
$\vec{\nu}^{\,*}$ and their 1D-marginalized posterior densities correspond excellently to the
priors densities. We further obtain $\chi^2 = 9.60$, which is a reduction compared to SM($\nu$-only)
by $3.80$. Given the now $5$ d.o.f., we find a $p$-value of $0.09$; the fit is therefore acceptable.
The evidence is $P(\text{data} \,|\, (9,10)) = (2.253 \pm 0.008) \cdot 10^{17}$. The most prominent
local mode lies within the rectangle
\begin{equation}
\begin{aligned}
    +3 & \leq \wilson{9}  \leq +9    &
    -6 & \leq \wilson{10} \leq -2\,,
\end{aligned}
\end{equation}
which contributes $(1.738 \pm 0.008) \cdot 10^{17}$ or roughly $77\%$ to the evidence.
The 1D marginal posteriors for both $\wilson{9}$ and $\wilson{10}$ are non-Gaussian,
and we find for their respective modes and $1\sigma$ intervals:
\begin{equation}
\begin{aligned}
    \wilson{9}   & = +5.9^{+0.7}_{-0.9}\,,  & \Delta_9    & = +1.6^{+0.7}_{-0.9}\,,\\
    \wilson{10}  & = -3.5^{+0.5}_{-0.8}\,,  & \Delta_{10} & = +0.7^{+0.5}_{-0.8}\,.
\end{aligned}
\end{equation}
In the above, we also state the ranges for $\Delta_i \equiv \wilson{i} -
\wilson{i}^\text{SM}$ for $i=9,10$. A comparison of our results with other findings
in the literature is difficult, due to the different methodologies. Naively, one
finds that the maximum distance between our results and the ones from
Refs.~\cite{Beaujean:2013soa,Altmannshofer:2014rta,Descotes-Genon:2015uva}
are
\begin{equation}
    -3.1\sigma \text{ for $\Delta_9$}\qquad\text{and}\qquad -1.1\sigma \text{ for $\Delta_{10}$},
\end{equation}
where we have expressed the distance in terms of the width of our results for $68\%$ probability
intervals.

In conclusion, we find that, while the (9,10) scenario can locally explain the data better by a
$\Delta \chi^2 = 3.80$, on average the SM($\nu$-only) scenario is more efficient in explaining
the data with posterior odds
\begin{equation}
    \frac{P((9,10)\,|\,\text{data})}{P(\text{SM($\nu$-only)}\,|\,\text{data})} = 1 : 6.6\,.
\end{equation}
Following Jeffreys' scale \cite{Jeffreys:1961} these odds are \emph{substantially} in favor of SM($\nu$-only).

\subsection{Scenario (9,10,9',10')}
\label{sec:scenario-SMp}

\begin{figure*}
\begin{tabular}{cc}
    \includegraphics[width=.4\textwidth]{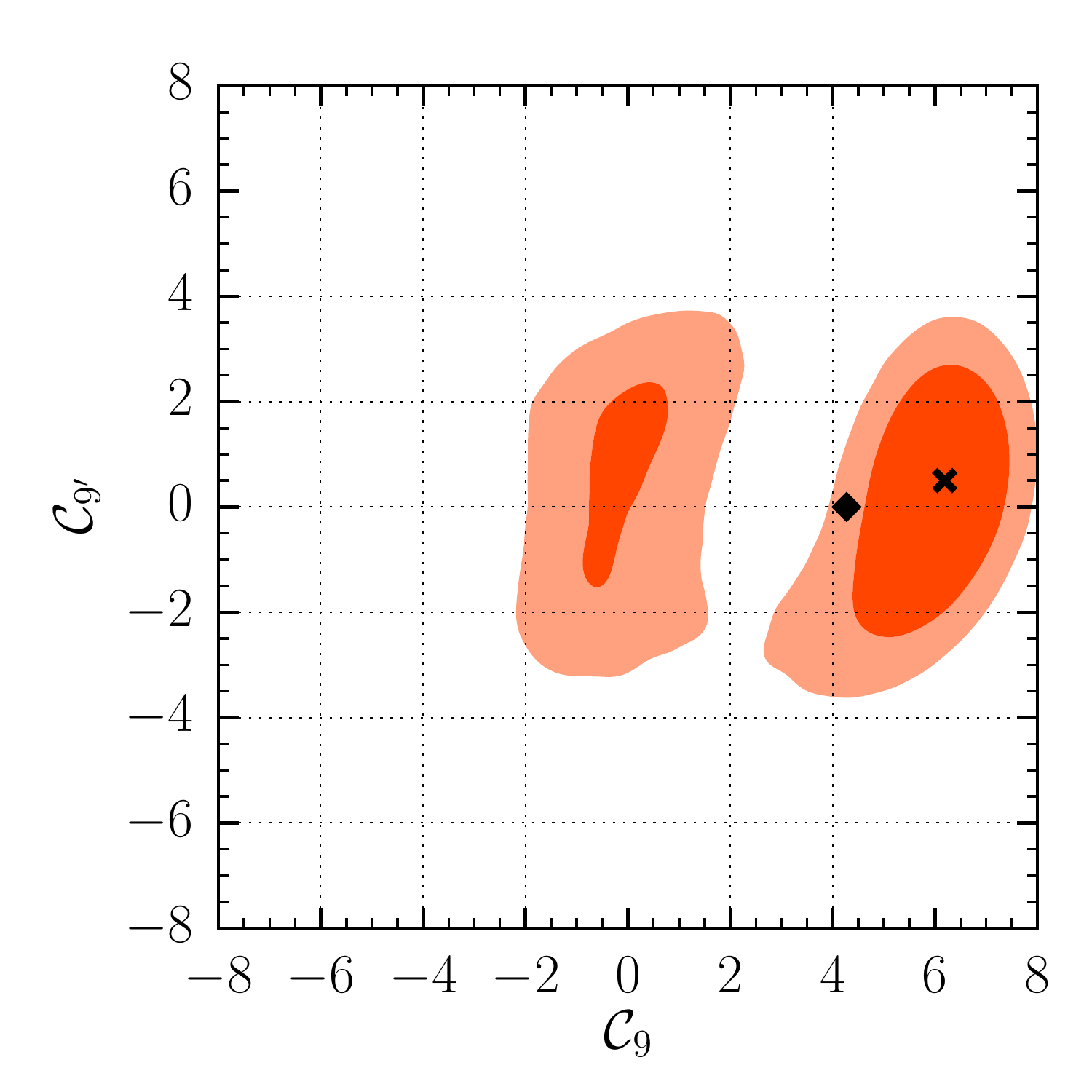} &
    \includegraphics[width=.4\textwidth]{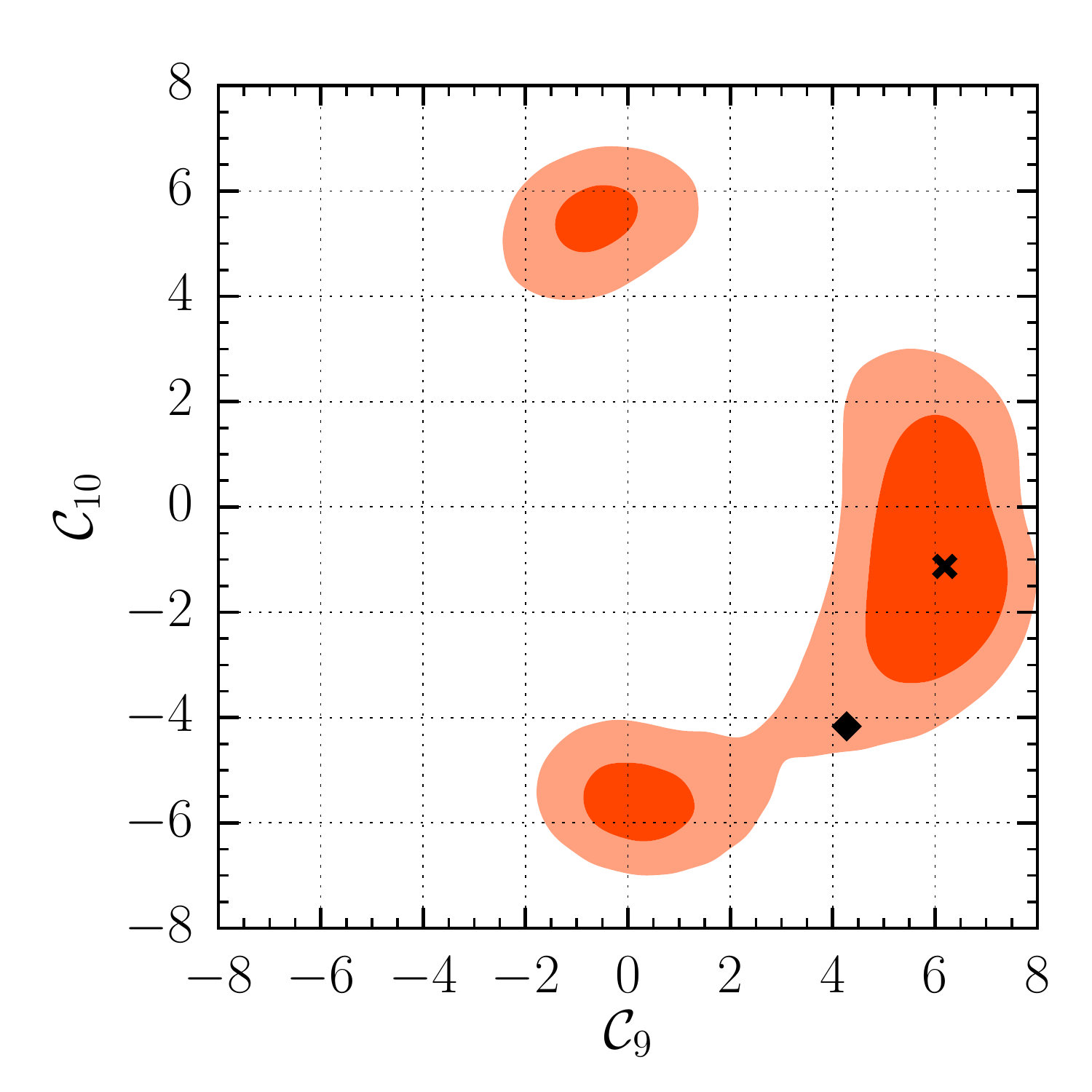}\\
    \includegraphics[width=.4\textwidth]{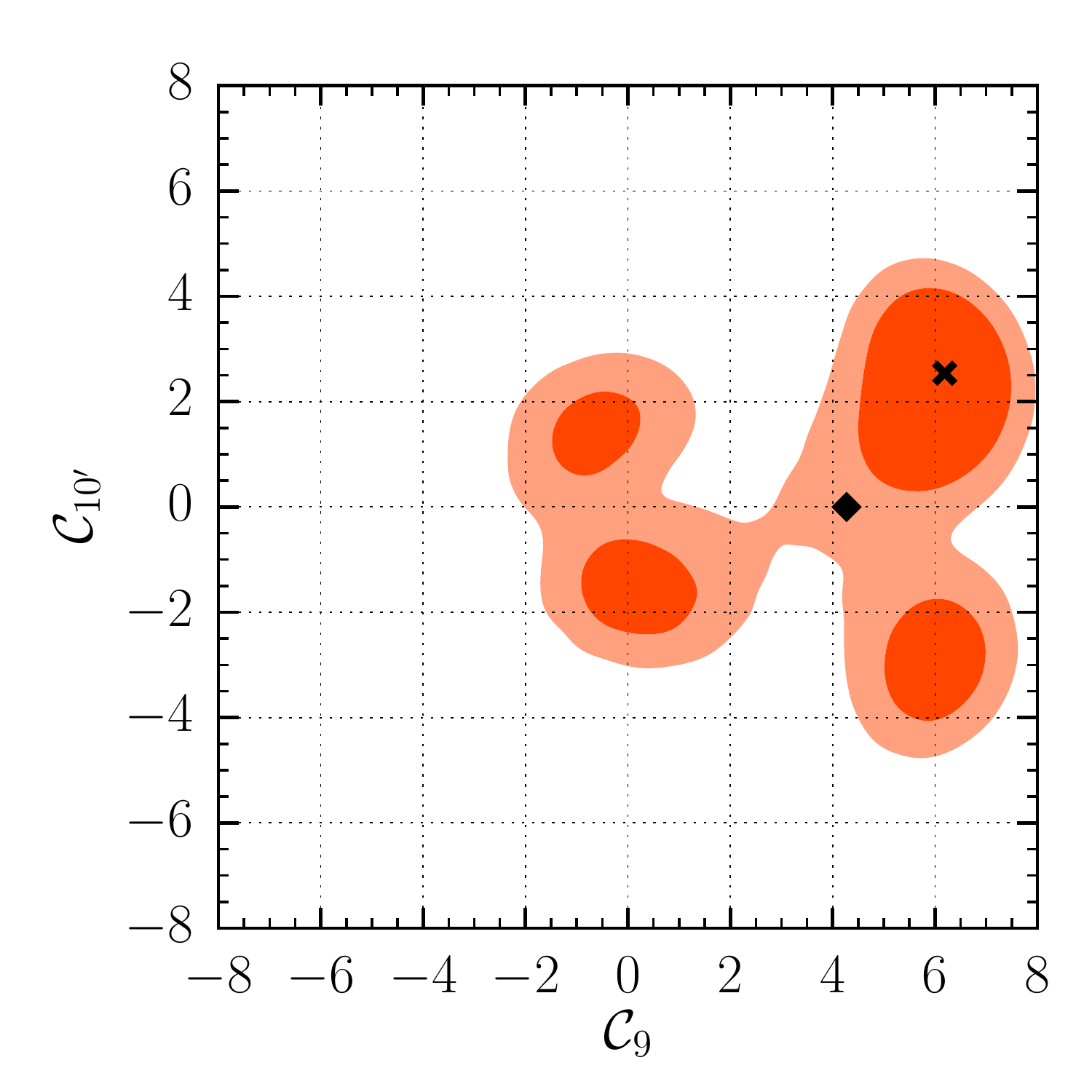} &
    \includegraphics[width=.4\textwidth]{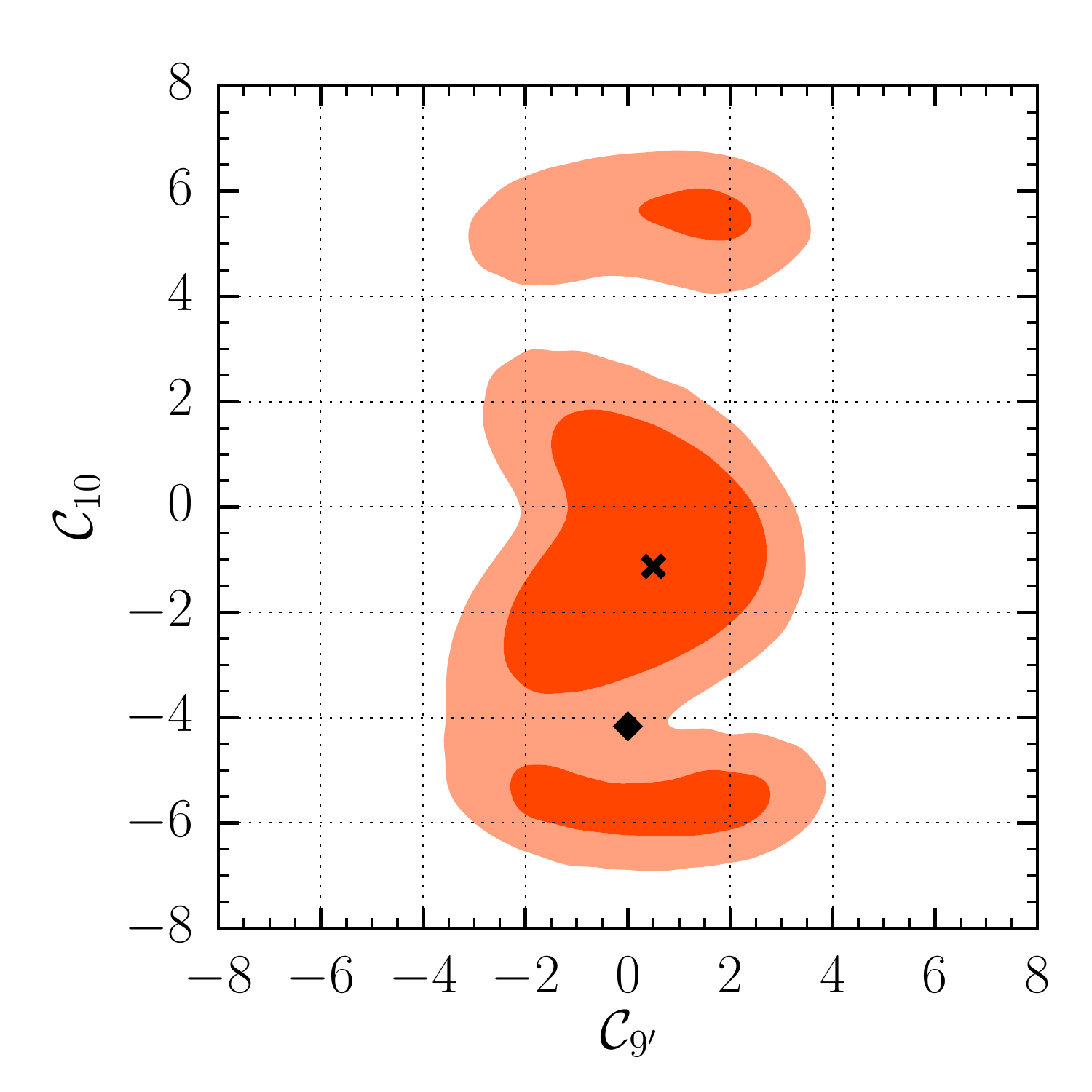} \\
    \includegraphics[width=.4\textwidth]{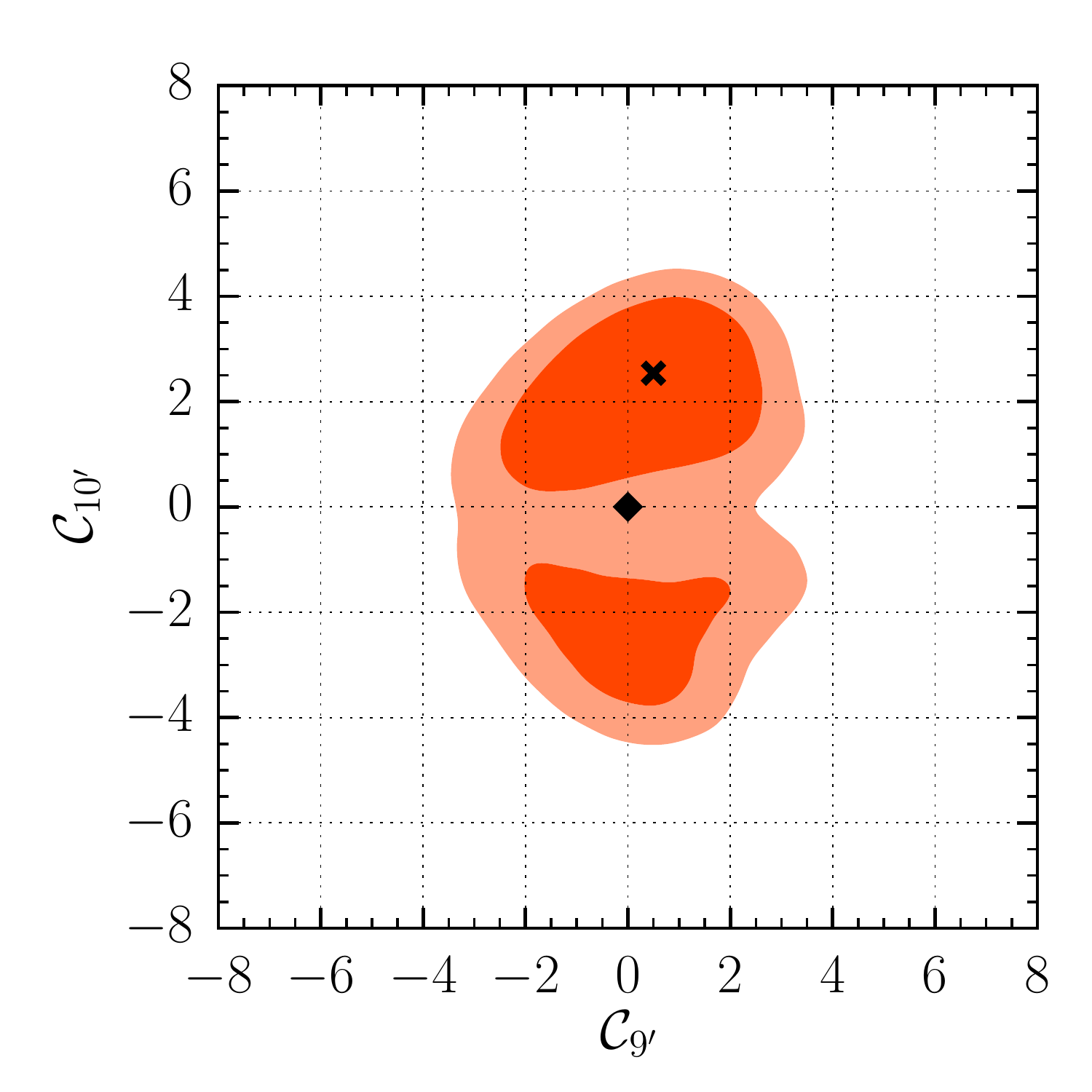} &
    \includegraphics[width=.4\textwidth]{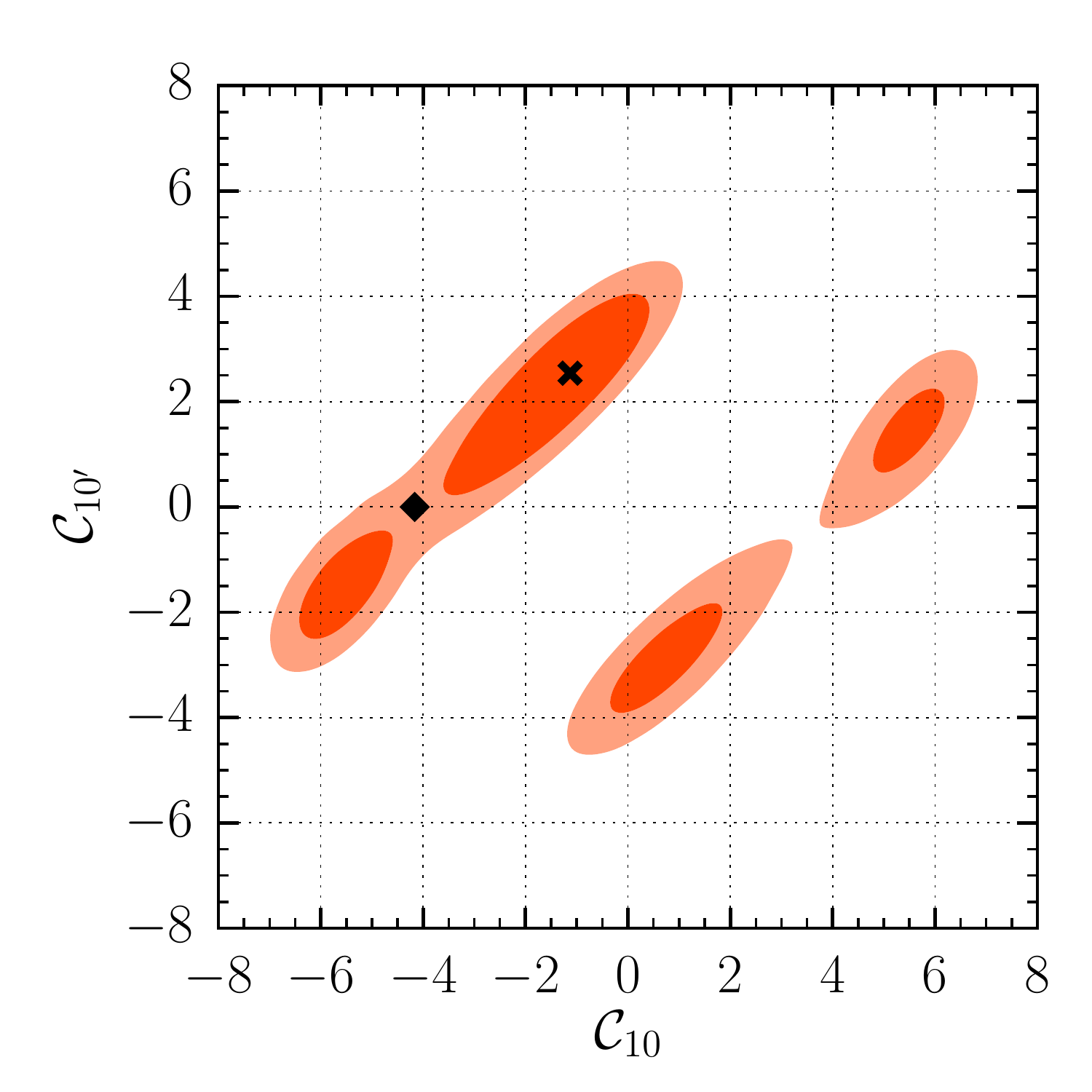}
\end{tabular}
\caption{\label{fig:posteriors-SMp}The 2D-marginalised posteriors for all pairs
    of Wilson coefficients in the $(9,10,9',10')$ scenario.
    The SM point is marked with a diamond shape, while the best-fit point from the full fit
    is marked with a black cross. The contours correspond to $68\%$ (inner contours)
    and $95\%$ (outer contours) of probability for the respective 2D-marginalised posteriors.
}
\end{figure*}

For this scenario we fit four parameters of interest in addition to the nuisance parameters,
yielding $39$ fit parameters. The $\theta$ components of the best-fit point are
\begin{equation}
    \vec{\theta}^{\,*}\!:\!(\wilson{9}, \wilson{9'}, \wilson{10}, \wilson{10'}) = ( 6.20, 0.50, -1.13, +2.54 )\,. \hspace{2ex}
\end{equation}
We show all 2D marginalizations of the posterior in \reffig{posteriors-SMp}. As
in Ref.~\cite{Beaujean:2013}, we find four local solutions. However, opposed to the
results of Ref.~\cite{Beaujean:2013}, the local solutions in our posterior are not
well separated in the 2D marginalizations. We interpret this as an effect
of a ``shallow'' posterior, which is due to the still considerable uncertainties
on the experimental results for the $\Lambda_b\to \Lambda$ observables, and
also the small number of observations: a full angular analysis of the decay
$\Lambda_b\to \Lambda(\to p\,\pi^-)\mu^+\mu^-$ is therefore desirable. As a further
consequence, the posterior of the nuisance parameters is very close to our
prior. At the best-fit point we find $\chi^2 = 3.87$, which is a reduction
compared to the $(9,10)$ scenario by $5.73$, and compared to SM($\nu$-only)
by $9.53$. Given the only 3 degrees of freedom in this exploratory scenario,
we obtain a $p$-value of $0.28$, which is a good fit. We obtain for the evidence
$P(\text{data}\,|\ (9,9',10,10')) = (2.188 \pm 0.003)\cdot 10^{16}$, with the uncertainty
only due to statistics.

We proceed to investigate one of the four solutions that is contained within
the hyperrectangle
\begin{equation}
\begin{aligned}
    +3 & \leq \wilson{9}   \leq +8\,, &
    -4 & \leq \wilson{9'}  \leq +4\,, \\
    -4 & \leq \wilson{10}  \leq +1\,, &
     0 & \leq \wilson{10'} \leq +5\,.
\end{aligned}
\end{equation}
Its local evidence is found to be $(1.152 \pm 0.001)\cdot 10^{16}$, which
corresponds to $\sim 53\%$ of the total evidence. We obtain the 1D marginalizations
within the above boundaries, which are non-Gaussian. The modes and
$1\sigma$ intervals read
\begin{equation}
\begin{aligned}
    \wilson{9}    & = +6.0^{+0.8}_{-0.8}\,, &    \Delta_9    & = +1.7^{+0.8}_{-0.8}\,, \\
    \wilson{9'}   & = +0.5^{+1.3}_{-1.8}\,,\\
    \wilson{10}   & = -1.3^{+1.3}_{-1.1}\,, &    \Delta_{10} & = +2.9^{+1.3}_{-1.1}\,,\\
    \wilson{10'}  & = +2.3^{+0.8}_{-1.3}\,,
\end{aligned}
\end{equation}
where, as before, $\Delta_i \equiv \wilson{i} -
\wilson{i}^\text{SM}$.

Our findings can be summarized as follows: The posterior odds
relative to the previous two fit scenarios are
\begin{equation}
    \frac{P((9,9',10,10')\,|\,\text{data})}{P(\text{SM($\nu$-only)}\,|\,\text{data})} = 1 : 100\,,
\end{equation}
as well as
\begin{equation}
    \frac{P((9,9',10,10')\,|\,\text{data})}{P((9,10)\,|\,\text{data})} = 1 : 15\,.
\end{equation}
Thus, again, SM($\nu$-only) is more efficient in its description of the data than
a new-physics interpretation involving $\wilson{9}$ through $\wilson{10'}$.

\section{Discussion}
\label{sec:summary}

The newly available lattice QCD results for the $\Lambda_b\to \Lambda$ form factors \cite{Detmold:2016pkz}
have considerably decreased the theoretical uncertainties in the $\Lambda_b\to \Lambda(\to p\,\pi^-) \mu^+\mu^-$ observables
at low hadronic recoil, and the strengths of the constraints on the $|\Delta B| = |\Delta S| = 1$ Wilson coefficients
are currently limited by the experimental uncertainties.
Nevertheless, already with the current experimental data \cite{Aaij:2015xza}, we find that this decay has now reached a similar level
of constraining power as the decay $\bar{B}\to \bar{K}^*\mu^+\mu^-$ exhibited after the first LHCb measurement \cite{Aaij:2011aa}.

Within our nominal fit in the $(9,10)$ scenario, we find
\begin{equation}
\begin{aligned}
    \wilson{9}   & = +5.9^{+0.7}_{-0.9}\,,  & \Delta_9    & = +1.6^{+0.7}_{-0.9}\,,\\
    \wilson{10}  & = -3.5^{+0.5}_{-0.8}\,,  & \Delta_{10} & = +0.7^{+0.5}_{-0.8}\,.
\end{aligned}
\end{equation}
Our fits in both the $(9,10)$ and $(9,9',10,10')$ scenarios were surprisingly well behaved, given the
small number of observables included. We look forward to including
the $\Lambda_b\to \Lambda(\to p\,\pi^-)\mu^+\mu^-$ data in a larger analysis
together with all mesonic decays.

Even though our fits of the Wilson coefficients yield noticeable reductions in $\chi^2$ compared to the SM,
neither the scenario $(9,10)$, nor the scenario $(9,9',10,10')$,
is as \emph{efficient} as the Standard Model in describing the combined present data on inclusive
$B\to X_s\ell^+\ell^-$ decays, the leptonic decay $B_s\to \mu^+\mu^-$,
and the branching ratio and angular observables of $\Lambda_b\to \Lambda(\to p\,\pi^-)\mu^+\mu^-$.
As a consequence, we find no evidence for effects of physics beyond the Standard Model.

When comparing our results for the Wilson coefficient $\wilson{9}$ with
analyses excluding the baryonic decay but including the decays $\bar{B}\to \bar{K}^{(*)}\mu^+\mu^-$, we find poor agreement
with the findings in the literature. The maximal distance emerges for the
most recent result of Ref.~\cite{Descotes-Genon:2015uva}, and reads $-3.1\sigma$
in terms of the standard deviation of our result.

In our opinion, the observed discrepancy can be caused by two different mechanisms:
\begin{enumerate}
    \item The discrepancy might arise from our incomplete understanding of the hadronic
        matrix elements of the two-point correlators of $\op{1,...,6;8}$ with the quark electromagnetic
        current, which effectively shift the Wilson coefficients $\wilson{7}$ and $\wilson{9}$.
        The main difficulty arises from the operators $\op{1}$ and $\op{2}$, whose contributions
        are enhanced by charmonium resonances (see e.g. Ref.~\cite{Khodjamirian:2010vf}, where
        these contributions are discussed within a hadronic dispersion relation).
        A drastically different shift to $\wilson{9}$ in the baryonic decay compared
        to the mesonic transitions, e.g. through different phases, would yield the different
        results that we currently face. This would constitute a breakdown
        of the universal structure of the transversity amplitudes at low
        recoil \cite{Bobeth:2010wg,Boer:2014kda} that is predicted by the OPE.
        We explicitly show in \refapp{lambdabonly} that such effects can only partially explain
        the presently observed shift to $\wilson{9}$.

    \item Given the large experimental uncertainties for the $\Lambda_b\to \Lambda(\to p\,\pi^-)\mu^+\mu^-$
        observables, statistical fluctuations could conspire to mimic a large
        positive shift to $\wilson{9}$. The best candidate for such an influence in
        the fit is the measurement of the branching ratio $\langle \mathcal{B}\rangle_{15,20}$.
        We note that the experimental uncertainty of $\langle \mathcal{B}\rangle_{15,20}$ \cite{Aaij:2015xza} is currently
        dominated by the uncertainty of the branching ratio of the normalization mode
        $\Lambda_b \to J/\psi\,\Lambda$ \cite{Agashe:2014kda}.
\end{enumerate}
One must also consider that the results of Ref.~\cite{Descotes-Genon:2015uva} are driven,
amongst other effects, by the low value of $R_K$ \cite{Aaij:2014ora}, which cannot be explained
by hadronic effects, and the consistent picture
of the mesonic decays $\bar{B}\to \bar{K}^{(*)}\mu^+\mu^-$ both below and above
the narrow charmonium resonances.

Ultimately, to settle the questions regarding $\wilson{9}$, we need both a reduction in the experimental uncertainties
for $\Lambda_b\to \Lambda(\to p\,\pi^-)\mu^+\mu^-$ (and an analysis of the full angular distribution,
e.g. using a principal moment analysis as proposed in \cite{Beaujean:2015xea})
and breakthroughs in our understanding of the nonlocal hadronic matrix elements of the operators $\op{1,...,6;8}$.

\acknowledgments

The work of S.M. is supported by National Science Foundation Grant Number PHY-1520996, and by the RHIC Physics Fellow Program of the RIKEN BNL Research Center.
The work of D.v.D. is supported by the Swiss National Science Foundation, grant PP00P2-144674.
We thank Christoph Bobeth, Joaquim Matias, Luca Silvestrini and Roman Zwicky for useful comments
on our preliminary results that were presented during the ``4th Workshop on Implications
of LHCb measurements and future prospects''.

\FloatBarrier

\appendix

\section{Parametrization of subleading terms in the low recoil OPE}
\label{app:subleading-terms}

The decay $\Lambda_b\to \Lambda\ell^+\ell^-$ can be described through
eight transversity amplitudes: $A_{\perp_0}^L$, $A_{\para_0}^L$, $A_{\perp_1}^L$, $A_{\para_1}^L$,
and their counterparts with $L \leftrightarrow R$. At low recoil, the OPE predicts a universal
structure, see \cite{Boer:2014kda}. Following Ref.~\cite{Beylich:2011aq}, this structure is broken
only by hadronic matrix elements $r_i$ (where $i\in\lbrace\perp_0,\para_0,\perp_1,\para_1\rbrace$)
at the level of dimension-five operators in the OPE. We therefore write the transversity amplitudes
as
\begin{widetext}
\begin{equation}
\begin{aligned}
    A_{\perp_0}^{L(R)}
        & = +\sqrt{2} N \sqrt{s_-} \frac{\mLambdaB + \mLambda}{\sqrt{q^2}}
        \left[
            C_{9,10,+}^{L(R)}\: f_0^V
            + \frac{2 m_b (C_7 + C_{7'})}{\mLambdaB + \mLambda} f_0^T\,\,
            + \left(\frac{4}{3} C_1 + C_2\right) r_{\perp_0}
        \right]\,,\\
    A_{\para_0}^{L(R)}
        & = -\sqrt{2} N \sqrt{s_+} \frac{\mLambdaB - \mLambda}{\sqrt{q^2}}
        \left[
            C_{9,10,-}^{L(R)}\: f_0^A
            + \frac{2 m_b (C_7 - C_{7'})}{\mLambdaB - \mLambda} f_0^{T5}
            + \left(\frac{4}{3} C_1 + C_2\right) r_{\para_0}\,
        \right]\,,\\
    A_{\perp_1}^{L(R)}
        & = -2 N \sqrt{s_-}
        \left[
            C_{9,10,+}^{L(R)}\: f_\perp^V
            + \frac{2 m_b (\mLambdaB + \mLambda)(C_7 + C_{7'})}{q^2} f_\perp^T\,\,\,
            + \left(\frac{4}{3} C_1 + C_2\right) r_{\perp_1}
        \right]\,,\\
    A_{\para_1}^{L(R)}
        & = +2 N \sqrt{s_+}
        \left[
            C_{9,10,+}^{L(R)}\: f_\perp^A
            + \frac{2 m_b (\mLambdaB - \mLambda)(C_7 - C_{7'})}{q^2} f_\perp^{T5}\,
            + \left(\frac{4}{3} C_1 + C_2\right) r_{\para_1}\,
        \right]\,,\\
\end{aligned}
\end{equation}
where the kinematics quantities $s_\pm$, the effective Wilson coefficients $C_{9,10,\pm}^{L(R)}$,
the normalization $N$, and the form factors $f_\lambda^J$ are defined as in Ref.~\cite{Boer:2014kda}.
\end{widetext}

In general, the matrix elements $r_i$ are complex-valued, $q^2$-dependent functions. These matrix elements arise only with a
suppression of order $\Lambda_\text{had}^2/Q^2$, where $Q^2 \sim \lbrace m_b^2, q^2\rbrace$.  (We note that a similar
parametrization is used in Ref.~\cite{Bobeth:2011gi}. However, there the OPE is used in terms of
Heavy-Quark Effective Theory (HQET)
operators \cite{Grinstein:2004vb}. As a consequence, the leading corrections to the OPE arise from
dimension-four operators, which enter suppressed by one power of the strong coupling $\alpha_s$.
As such, their effect is virtually the same as here.)

Since in our fits we only use a single $q^2$ bin that covers the entire phase space above $q^2 = 15\,\GeV^2$,
we can parametrize the unknown hadronic matrix elements as $q^2$-constant quantities.
Within the fits, we take the $r_i$ to be real-valued\footnote{%
    Only the observables $K_{3sc}$ and $K_{3s}$, or combinations thereof, are sensitive to the
    phases of the matrix elements $r_i$. As these observables are presently unconstrained, our
    use of real-valued quantities therefore suffices. A similar observation has been made in
    \cite{Beaujean:2013soa} for power corrections at large $q^2$ in $B\to K^{(*)}\mu^+\mu^-$ decays.
}, with \emph{uncorrelated} Gaussian priors centered around zero and with a standard
deviation of $0.03$: $r_i \sim \mathcal{N}(0, 0.03)$. We emphasize that this is a conservative estimate
for the size of these hadronic matrix elements, since $\Lambda_\text{had}^2/Q^2 \lesssim 0.9\%$.

\section{Posterior-predictive distributions for the $\Lambda_b\to \Lambda(\to p\,\pi^-)\mu^+\mu^-$ angular observables}
\label{app:predictions}

In this section, we compute posterior-predictive distributions for the normalized angular observables
\begin{equation}
    \hat{K}_n \equiv \frac{\langle K_n\rangle_{15,20}}{\langle \Gamma\rangle_{15,20}}\,,
\end{equation}
from the $(9,10)$ fit scenario.
Summaries in the form of the mode and the $68\%$ probability interval for the
observables with $n \in \lbrace 1ss, 1cc, 1c, 2ss, 2cc, 2c, 4sc, 4s\rbrace$ are
shown in \reftab{predictions}.  We abstain from providing predictions for the
observables with $n \in \lbrace 3sc, 3s\rbrace$, since for real-valued Wilson
coefficients these observables are only sensitive to small interference
effects introduced by the imaginary parts of the hadronic matrix elements of
$\op{1,...,6;8}$, and by the contributions proportional to $V_{ub}$.

\begin{table}
\renewcommand{\arraystretch}{1.4}
\newcolumntype{C}[1]{>{\centering\let\newline\\\arraybackslash\hspace{0pt}}m{#1}}
\begin{tabular}{c|C{.11\textwidth}}
    Observable          & $(9,10)$                   \\
    \hline
    $\hat{K}_{1ss}$     & $+0.352^{+0.003}_{-0.003}$ \\
    $\hat{K}_{1cc}$     & $+0.296^{+0.006}_{-0.006}$ \\
    $\hat{K}_{1c}$      & $-0.233^{+0.008}_{-0.008}$ \\
    \hline
    $\hat{K}_{2ss}$     & $-0.195^{+0.005}_{-0.005}$ \\
    $\hat{K}_{2cc}$     & $-0.153^{+0.006}_{-0.006}$ \\
    $\hat{K}_{2c}$      & $+0.186^{+0.004}_{-0.004}$ \\
    \hline
    $\hat{K}_{4sc}$     & $-0.022^{+0.005}_{-0.005}$ \\
    $\hat{K}_{4s}$      & $-0.102^{+0.007}_{-0.009}$ \\
    \hline
\end{tabular}
\renewcommand{\arraystretch}{1.0}
\caption{
    Summary of the 1D marginalised posterior-predictive distributions for the
    normalized angular observables $\hat{K}_n = \langle K_n\rangle_{15,20} / \langle \Gamma \rangle_{15,20}$.
    We present the distributions obtained from the posteriors of the $(9,10)$
    scenario. The statistical uncertainty from the Monte Carlo
    integration is estimated to be $10^{-3}$.
}
\label{tab:predictions}
\end{table}

\section{Fits of $\Lambda_b\to \Lambda(\to p\,\pi^-)\mu^+\mu^-$ data only}
\label{app:lambdabonly}

In order to further investigate a possible hadronic origin for the tensions between theory and experiment,
we carry out fits to only the $\Lambda_b\to \Lambda(\to p\,\pi^-)\mu^+\mu^-$ 
observables listed in \reftab{goodness-of-fit}.
Beside the scenario SM($\nu$-only), we also employ a new scenario $(9)$:
\begin{equation}
    (9) : \begin{cases}
        \wilson{9}              & \in [-4,+9]\\
        \wilson{7,7',9',10,10'} & \text{SM values}\\
        \vec\nu                 & \text{free floating}
    \end{cases}\,.\\
\end{equation}
Using the very
same priors as described in \refsec{framework}, we find poor fits: When only
fitting the nuisance parameters, the $p$-value is $1.3\cdot 10^{-2}$, while
the fit with freely floating $\wilson{9}$ only slightly improves the $p$-value
to $1.5\cdot 10^{-2}$. The reason for this behavior is that the current experimental
results for the observables $\la \mathcal{B} \ra_{15,20}$ and $\la A_\text{FB}^\ell\ra_{15,20}$ pull
$\wilson{9}$ in opposite directions (the branching ratio $\la \mathcal{B} \ra_{15,20}$, which more strongly depends on
$\wilson{9}$, favors a positive shift).

We thus investigate the
possibility that hadronic effects break the universal nature of the OPE
for the transversity amplitudes in $\Lambda_b\to \Lambda\mu^+\mu^-$ transitions.
This corresponds to a breakdown of the semi-local quark-hadron duality.
We can simulate such effects by dramatically increasing the allowed ranges of the power corrections $r_i$ as
defined in \refapp{subleading-terms}. We let
$r_i \sim \mathcal{N}(0, 3)$, on a support $-5 \leq r_i \leq +5$. Using
these priors, we repeat our fits to only the $\Lambda_b\to\Lambda(\to p\,\pi^-)\mu^+\mu^-$ observables.

For the SM($\nu$-only) scenario with the wide priors for $r_i$ we obtain $\chi^2 = 4.27$, and a $p$-value of
$0.37$. All 1D-posteriors of the form factor parameters, CKM parameters and
quark masses are in agreement with their priors. The posteriors for the power
correction parameters can be summarized as follows:
\begin{equation}
\begin{aligned}
    r_{\perp,0} & =  +2.1^{+2.4}_{-2.2}\,, \\
    r_{\para,0} & =  +3.6^{+1.4}_{-1.8}\,, \\
    r_{\perp,1} & =  -3.2^{+2.1}_{-1.8}\,, \\
    r_{\para,1} & =  -1.5^{+2.2}_{-2.5}\,. \\
\end{aligned}
\end{equation}
The total evidence is $P(\text{only $\Lambda_b\to \Lambda\mu^+\mu^-$}\, |\, \text{SM($\nu$-only)}) = 1.8\cdot 10^5$.

For the scenario $(9)$ with the wide priors for $r_i$ we obtain $\chi^2 = 4.59$, and a $p$-value of $0.20$.
This is surprising at first glance, since it means that adding one parameter
has lead to an \emph{increase} in $\chi^2$. However, $-\log P(\vec{x}^*\,|\, (9),
\text{only $\Lambda_b\to \Lambda\mu^+\mu^-$})$ has in fact \emph{decreased} by
$4.35$ on the log scale. With respect to SM($\nu$-only), the $r_i$ components
of the best-fit point have moved closer to $0$, thereby increasing the
posterior. At the same time, a shift of $\Delta_9 = 0.7^{+1.5}_{-1.3}$
compensates for the smaller values of the parameters $r_i$.
The 1D posteriors of the power corrections read:
\begin{equation}
\begin{aligned}
    r_{\perp,0} & =  +1.8^{+2.4}_{-2.5}\,, \\
    r_{\para,0} & =  +2.9^{+1.9}_{-1.9}\,, \\
    r_{\perp,1} & =  -3.4^{+2.4}_{-1.6}\,, \\
    r_{\para,1} & =  -2.5^{+2.3}_{-2.0}\,. \\
\end{aligned}
\end{equation}

Compared to the shift $\Delta_9\Big|_{(9,10)} = 1.6^{+0.7}_{-0.9}$, we
see a marked reduction in the need to modify $\wilson{9}$.
We conclude that symmetry-breaking shifts to all four transversity amplitudes
in $\Lambda_b\to \Lambda\mu^+\mu^-$ can only partially explain our results
in the $(9,10)$ scenario.

\vspace{1ex}

\bibliography{references}

\end{document}